\documentclass[12pt]{iopart}

\usepackage{iopams}  
\newcommand{\be}{\begin{equation}}
\newcommand{\ee}{\end{equation}}
\newcommand{\beqn}{\begin{equation}}
\newcommand{\eeqn}{\end{equation}}
\newcommand{\bea}{\begin{eqnarray}}
\newcommand{\eea}{\end{eqnarray}}

\begin{document}

\title[Nuclear single-particle states]{Nuclear single-particle 
states: dynamical shell model and energy density functional 
methods}

\author{P.F. Bortignon$^{1,2}$, G. Col\`o$^{1,2}$, and 
H. Sagawa$^{3}$}

\address{$^{1}$ Dipartimento di Fisica, Universit\`a
                degli Studi di Milano, via Celoria 16, Milano, Italy}
\address{$^{2}$ Istituto Nazionale di Fisica Nucleare (INFN), Sez. di
                Milano, via Celoria 16, Milano, Italy}
\address{$^{3}$ Center for Mathematics and Physics, University of Aizu,
                Aizu-Wakamatsu, Fukushima 965-8560, Japan}

\ead{bortignon@mi.infn.it, colo@mi.infn.it, sagawa@u-aizu.ac.jp}

\begin{abstract}
We discuss different approaches to the problem of
reproducing the observed features of nuclear single-particle
(s.p.) spectra. In particular, we analyze the 
dominant energy peaks, and the single-particle strength 
fragmentation, using the example of neutron states in
$^{208}$Pb. Our main emphasis is the interpretation of that
fragmentation as due to particle-vibration coupling (PVC). 
We compare with recent Energy Density Functional 
(EDF) approaches, and try to present a critical perspective.
\end{abstract}

\section{Introduction}

One of the most striking features of nuclei is the validity
of the shell model where the nucleons move independently in a smooth
single-particle (s.p.) potential. However, self-consistent
mean field (SCMF) approaches \cite{Heenen} are much more sucessful in
reproducing the bulk properties that emerge from the 
mean field picture (like masses, that can nowadays be
reproduced by Skyrme-HFB calculations with a remarkable
accuracy of about 0.6 MeV \cite{Goriely}) than
in accounting for the s.p. states. 

If SCMF is viewed as an approximate realization of
an exact, yet unknown and perhaps unfeasible, Density
Functional Theory (DFT), it is not surprising that
quantities like the total energy are well accounted
for since they belong to the domain of this theory
based on the well-known Kohn-Sham theorem. (We do not
enter here the problem if this theorem can be
expected to hold in a self-bound system since this
discussion can be found elsewhere in the present 
volume). 

If single-particle states are not, strictly speaking, 
within the DFT framework, two possible
routes can be undertaken. Some authors are currently
trying to improve the accuracy of present DFT implementations
aiming at functionals with so-called ``spectroscopic''
accuracy (referring to, of course, single-particle
spectroscopy). Another point of view, which is
the preferred one by the authors of the present
contribution, is to generalize the shell model to
the so-called ``dynamical'' shell model. It has to be
noted that the former attitude does not solve the
problem of evaluating the so-called spectroscopic
factors or, more generally, the well-known
fragmentation of the s.p. strength.

The potential emerging from SCMF calculations is 
static, but in general nonlocal in space (or, in 
equivalent words, velocity- or momentum-dependent).
The fluctuations of this average potential lead to 
collective modes, in particular surface vibrations.
Taking into account the coupling of these modes to 
the s.p. motion, the shell model acquires a dynamical 
content. Thus, the average potential becomes also nonlocal 
in time, being characterized by an energy (or frequency) 
dependence.

The velocity dependence of the s.p. potential can be 
characterized by the so-called $k$-mass $m_k$, and its 
frequency dependence by the $\omega$-mass $m_\omega$. 
Most experiments only probe the product of these two 
quantities, that is, the ``total'' effective mass $m^*$
(see the definitions in Sec. \ref{formalism}). 
The dynamics of the shell model affects different 
nuclear properties as: the fragmentation and related 
spectroscopic factors of the s.p. states, their
density (which is proportional to $m^*$ near the Fermi 
energy), the s.p. spreading widths, and the imaginary 
component of the optical potential. 
A unified description of the s.p. motion
at positive and negative energy eventually emerges.

The importance of the dynamical couplings was first 
demonstrated in the case of electrons in metals and 
of normal liquid $^3$He. These results inspired
the work of Bertsch and Kuo \cite{BK} on the enhancement 
of the effective mass near the Fermi energy in finite 
nuclei. The review article \cite{CM} contains the 
formalism reported in the next Section as well as the many
results obtained in the 80s. The expression ``dynamics of 
the shell model'' is after G.E. Brown \cite{GB}. 

The old calculations of the 80s are mostly not
consistent. It is hard to extract quantitative conclusions
because of the many approximations involved. More
consistent calculations have been recently done. They
can be confronted with modern EDF-based approaches. We
shall do it in the simple case of $^{208}$Pb, showing
that the problem of the s.p. states is still an
open one.

\section{Formalism of the dynamical shell model}
\label{formalism}

One can introduce using standard many-body techniques a  
``mass operator'' $\cal M$, defined by  
\be
\left[ \omega + (2m)^{-1} \nabla^2_r \right] G(\vec r, 
\vec r^\prime; \omega) = \delta(\vec r-\vec r^\prime) + 
\int d_3r^\prime\ {\cal M}(\vec r,\vec r^\prime; \omega) 
G(\vec r,\vec r^\prime; \omega)
\label{eq:1} 
\ee
in terms of the one-body Green function (Lehmann representation)
\be
G(\vec r,\vec r^\prime; \omega) = \sum_h \frac{\phi_h(\vec r)
\phi^*_h(\vec r^\prime)}{\omega-\omega_h-i\eta} +
\sum_p \frac{\phi_p(\vec r)
\phi^*_p(\vec r^\prime)}{\omega-\omega_p+i\eta}.
\label{eq:2}
\ee
Here 
\bea
\phi_h(\vec r) = \langle \psi_h^{(A-1)} \vert a(\vec r) \vert 
\psi_0^{(A)} \rangle, \nonumber \\
\phi_p(\vec r) = \langle \psi_0^{(A)} \vert a(\vec r) \vert
\psi_p^{(A+1)} \rangle.
\eea
$\psi_h^{(A-1)}$ is a suitably normalized eigenstate of
the Hamiltonian of the (A-1)-particle system with energy 
$E_h^{(A-1)}$, $\psi_p^{(A+1)}$ is a suitably normalized eigenstate of
the Hamiltonian of the (A+1)-particle system with energy
$E_p^{(A+1)}$, and
\be\label{eq:4}
\omega_p = E_p^{(A+1)} - E_0^{(A)} \ \ \ \ \ 
\omega_h = E_0^{(A)} - E_h^{(A-1)}.
\ee  
Eqs. (\ref{eq:1}) and (\ref{eq:2}) give 
\be\label{we}
\left[ \omega + (2m)^{-1} \nabla^2 \right] \phi_p(\vec r) 
- \int d_3r^\prime\ {\cal M}(\vec r,\vec r^\prime; \omega_p) 
\phi_p(\vec r^\prime) = 0. 
\ee
This s.p. wave equation has two important properties:
i) It has discrete eigenvalues corresponding to the 
energies of the bound states of the (A+1) and (A-1) systems, 
ii) In the continuum gives the exact diagonal (elastic scattering) 
element of the S matrix \cite{BS}. 

The mass operator, in what follows, will contain the 
Hartree-Fock (HF) mean field and the contribution of the 
coupling of the s.p. HF states to the vibrational 
modes (particle-vibration coupling or PVC), in particular 
to collective modes obtained from self-consistent 
linear response theory (Random Phase Approximation or 
RPA). 

Although one could solve (\ref{we}) directly, the usual
way has been so far to solve it at the HF level and add
the contribution of the particle-vibration part of the
mass operator within second-order perturbation theory. 
The expressions for this second-order correction can be
found in most of the references quoted herein. 

To work out the expressions for $m_k$, $m_\omega$ and $m^*$, 
it is easier to consider a uniform system \cite{CM}. Thus, 
having the relation
\be
E = \frac{k^2}{2m} + {\cal V}(k;E)
\ee
between the energy $E$ and the momentum $k$ of the 
quasiparticle (dressed s.p.), and
${\cal V}$ being the real part of the complex mass operator, 
the effective mass is defined
\be
\frac{dE}{dk} = \frac{k}{m^*}.
\ee
Then, it is easy to write
\bea
\frac{m_k}{m} & = & \left\{ 1 + \frac{m}{k} 
\left[ \frac{\partial{\cal V}(k;\omega)}{\partial k} 
\right] \right\}^{-1}_{\omega=E(k)}, \nonumber \\
\frac{m_\omega}{m} & = & 1- \left[  
\frac{\partial}{\partial\omega} {\cal V}(k;\omega) 
\right]_{\omega=E(k)},
\eea
to obtain
\be
\frac{m^*}{m} = \frac{m_k}{m}\cdot\frac{m_\omega}{m}. 
\ee

\section{The contradicting results}

\subsection{Particle-vibration coupling: old and new results}

In Tables 4.3a and 4.3b of Ref. \cite{CM}, an extensive review 
of the results obtained for $m_\omega$ in $^{208}$Pb by nine 
groups in the period 1968-1983 can be found. Rather different
frameworks had been adopted, no one being fully self-consistent: 
s.p. potentials range from harmonic oscillator (HO) to Woods-Saxon 
(WS) or HF with Skyrme forces; residual interactions at the
particle-vibration vertex are either multipole-multipole forces, or forces
of Landau-Migdal type, or Skyrme forces but with velocity-dependent terms 
dropped, or even G-matrix interactions. 

The average value for $m_\omega$ is in the 
range 1.2-1.4 for the neutron valence shells, and in
the range 1.3-1.6 for the proton case. The inverse of $m_\omega$ 
gives the spectroscopic factor reported
in table 4.4 of Ref. \cite{CM}. In general,
it has been impossible to find an accurate agreement
between theory and experiment, either at the level
of main energy centroids of the s.p. strength
distribution, or at the level of spectroscopic factors. 
The problem of spectroscopic factors had been not
considered too seriously, in view of the large experimental
errors and ambiguities associated with them. Nowadays, new analysis
are being carried out \cite{Tsang}, but the theoretical 
ambiguities associated with
the extraction of spectroscopic factors are still
under discussion (see e.g. the contribution by 
R.~J. Furnstahl and A. Schwenck in the present volume). 

If we concentrate on s.p. energies, the most consistent
among the calculations reported in \cite{CM}, namely that
by V. Bernard and N. Van Giai \cite{Bernard}, still underestimates
the density of levels.

\begin{table}[htbp]
\caption{\label{shifts} Shifts (that, is differences between
dressed and bare single-particle energies) associated with
the dominant s.p. neutron components in $^{208}$Pb. See the
text for a discussion of how they have been obtained.}
\small\rm
\begin{tabular}{|c|rrr|}
\hline
& \multicolumn{3}{|c|}{$\Delta\varepsilon=
\varepsilon_i-\varepsilon^{(0)}_i$} \\
& \cite{PR} & \cite{Dubrovnik} & \cite{Zalewski} \\
\hline
3d$_{3/2}$  & -0.61  & -0.32  &  0.09  \\
4s$_{1/2}$  & -0.56  & -0.21  &  0.24  \\
3d$_{5/2}$  & -0.76  & -0.43  &  0.07  \\
1j$_{15/2}$ & -1.36  & -0.55  &        \\
2g$_{9/2}$  & -0.79  & -0.40  &  0.03  \\
1i$_{11/2}$ & -0.31  & -0.37  & -0.14  \\
\hline
3p$_{1/2}$  & -0.02  &  0.01  &  0.05  \\
2f$_{5/2}$  &  0.33  &  0.01  &  0.12  \\
3p$_{3/2}$  &  0.14  &  0.03  &  0.06  \\
1i$_{13/2}$ &  0.49  &  0.05  &  0.22  \\
1h$_{9/2}$  &  1.40  &  0.06  &        \\
2f$_{7/2}$  &  1.40  &  0.63  &  0.12  \\
\hline
\end{tabular}
\end{table}

The most recent calculations for $^{208}$Pb in 
the framework of dynamical shell model, or PVC model, 
have been reported in Refs. \cite{PR} and \cite{Dubrovnik}.
The results are displayed in Table \ref{shifts} 
and \ref{energies} and compared with the 
results of Ref. \cite{Zalewski} (cf. below) and 
with the experimental energies, in the case of the 
neutron valence particle and hole states. 
In Table \ref{shifts} we display, for Refs. 
\cite{PR} and \cite{Dubrovnik}, the differences
between the s.p. energies $\varepsilon_i$ obtained
with PVC included, and the mean field energies 
$\varepsilon^{(0)}_i$. These differences 
$\Delta\varepsilon_i$ will be called ``shifts'' 
in what follows. (See below for a definition of
the shifts from Ref. \cite{Zalewski}). 

In the covariant theory of Ref. \cite{PR}, the mean
field is associated with the NL3 \cite{NL3} effective
Langrangian, which is also employed to obtain the 
phonons within the relativistic RPA (RRPA). The Dyson 
equation (\ref{eq:2}) for the s.p. Green's functions 
is solved in the diagonal approximation for the 
mass operator $\cal M$. A rather simple 
particle-vibration model is used, in which the 
coupling vertex is obtained with the matrix element 
of the residual interaction times the transition 
density of the vibrations.  

The calculations of Ref. \cite{Dubrovnik} are performed 
in the framework of Skyrme HF, with the PVC contributions 
to the mass operator $\cal M$ calculated within
second-order perturbation theory. The phonons are
obtained through fully self-consistent RPA, and all 
the terms of the particle-hole (p-h) interaction 
are kept also in the PVC vertex. There are strong 
cancellations due to the velocity dependent terms 
that lead to small shifts, especially for the hole states.
The calculations have been done using the 
SLy5 Skyrme set \cite{Chabanat}. Calculations with 
other Skyrme functionals are in progress. 

\begin{table}[htbp]
\caption{\label{energies} Dressed single-particle 
energies and spectroscopic factors $S$ 
associated with the dominant s.p. neutron 
components in $^{208}$Pb, compared with the 
experimental findings.}
\small\rm
\begin{tabular}{|c|rrr|r|rr|l|}
\hline
& \multicolumn{3}{|c|}{$\varepsilon_{\rm th}$} & 
$\varepsilon_{\rm exp}$ & 
\multicolumn{2}{|c|}{$S_{\rm th}$} & 
$S_{\rm exp}$ \\
& \cite{PR} & \cite{Dubrovnik} & \cite{Zalewski} & 
& \cite{PR} & \cite{Dubrovnik} & \\
\hline
3d$_{3/2}$  & -0.63  &  0.01  &  0.78  & -1.40 
            &  0.89  &  0.94  &  0.88$\pm$0.1  \\
4s$_{1/2}$  & -0.92  & -0.31  &  0.80  & -1.90 
            &  0.92  &  0.95  &  0.88$\pm$0.1  \\
3d$_{5/2}$  & -1.39  & -1.08  & -0.43  & -2.37 
            &  0.88  &  0.91  &  0.88$\pm$0.1  \\
1j$_{15/2}$ & -1.84  & -1.75  &        & -2.51 
            &  0.66  &  0.86  &  0.53$\pm$0.1  \\
2g$_{9/2}$  & -3.29  & -3.62  & -3.16  & -3.94 
            &  0.84  &  0.93  &  0.78$\pm$0.1  \\
1i$_{11/2}$ & -3.28  & -1.38  & -1.67  & -3.16 
            &  0.88  &  0.94  &  0.96$\pm$0.2  \\
\hline
3p$_{1/2}$  & -7.68  & -8.04  & -8.06  & -7.37 
            &  0.90  &  0.93  &  1.07 \\
2f$_{5/2}$  & -8.77  & -8.94  & -8.91  & -7.94 
            &  0.86  &  0.94  &  1.13 \\
3p$_{3/2}$  & -8.27  & -9.16  & -9.17  & -8.27 
            &  0.86  &  0.91  &  1.00 \\
1i$_{13/2}$ & -9.11  & -10.13 & -9.30  & -9.00 
            &  0.81  &  0.94  &  1.04 \\
1h$_{9/2}$  & -11.96 & -12.01 &        & -10.78 
            &  0.36  &  0.94  &  1.10 \\
2f$_{7/2}$  & -9.71  & -11.44 & -11.90 & -9.71 
            &  0.64  &  0.75  &  0.88 \\
\hline
\end{tabular}
\end{table}

In Ref. \cite{Zalewski}, the HF problem is solved 
for the even-even and even-odd (A$\pm$1) systems 
using the SLy4$_L$ interaction. The ``dressed'' 
energies $\varepsilon_i$ are obtained 
from an equation like the present Eq. (\ref{eq:4}). 
In this case $E_0$ is the total energy of the
even-even system and $E_p$ ($E_h$) is the total
energy of the odd system when a given occupancy
(vacancy) is imposed. The ``bare'' energies are
obtained by excluding the so-called mass, shape 
and spin polarization effects: in practice this
corresponds, respectively, to neglecting the proper
change of the center-of-mass correction, to 
imposing spherical symmetry, and to discarding 
time-odd terms. 

One can see from Table \ref{shifts} that shifts 
obtained by RMF plus PVC and Skyrme plus PVC are
different, and it is hard to connect either of them
to the calculations of \cite{Zalewski}. If we
look now at the final ``dressed'' energies obtained
in the different approaches (Table \ref{energies}), 
and compare with
experimental values, we conclude that the open 
problem is that even for the ``benchmark'' nucleus 
$^{208}$Pb we do not dispose of a consistent 
calculation which reproduces well the experimental 
data. This is also true for spectroscopic factors. 

In the calculations of the spectroscopic factors, 
the effects of the short-range correlations (SRC) 
should also be included. On the one side, in 
a recent work \cite{Barbieri}, it is confirmed 
that the PVC involving long-range correlations (LRC) 
can be identified as the main mechanism, the 
SRC explaning only a small fraction (up to about 10\%) 
of the deviations of the spectroscopic factors 
from the independent s.p. values. Consistently, 
from a qualitative point of view, in Ref. \cite{CM} 
it was concluded that the main contribution of 
highly excited core polarized states to $m_\omega$, 
and therefore to the spectroscopic factors, 
can be considered as a constant in the vicinity 
of the Fermi energy. The subtracted dispersion relation
technique, with different choices for the behaviour 
of the imaginary part of the mass operator at high 
energy, was used.

\subsection{EDF extensions}

The lack of an accurate description of s.p. states
is also clear for EDF practitioners not only in 
keeping with the findings of Ref. \cite{Zalewski}. 
The authors of \cite{Kortelainen} have concluded that standard
forms of the Skyrme energy density functional do not 
allow that accurate description. The effect of tensor
forces has been studied very carefully by the
authors of \cite{Lesinski1,Lesinski2}, but their 
conclusion is that the tensor terms, albeit important, cannot
remedy the deficiencies of the central terms: in
practice, they cannot produce a functional which is
more accurate as far as s.p. states are concerned. 

In Ref. \cite{Carlsson}, gradient corrections up to
next-to-next-to-next-to-leading order (N$^3$LO, in practice
derivatives up to sixth power) have been added to a 
local EDF, in order 
to better reproduce surface effects, for example on the effective 
mass $m_\omega$ (cf. also \cite{Zalewski1}).
The precision on s.p. states is improving \cite{Carlsson_prep}.

\section{Conclusions} 

We still lack, despite the tremendous progress in many
aspects of nuclear structure physics, an accurate
description of s.p. strength distributions. While
new radioactive beam facilities have already started
to show how this kind of observables evolve with 
the decrease of 
the neutron or proton separation energies \cite{Gade1,Gade2}, 
even for standard nuclei we have troubles to get
accuracy for dominant s.p. energies and
spectroscopic factors (using models that have had
big success in reproducing other kind of observables). 

The addition of higher power gradient corrections 
to a local EDF may improve the s.p. energies. We deem, 
however, that a dynamical theory is necessary to describe 
quantities like the spectroscopic factors.
The merging of the two approaches defines anyway, 
to some extent, the future of this fundamental 
field of nuclear structure research. In a many-body
approach like the dynamical shell model one believes
that the basic object should be a time- or 
energy-dependent quantity like the s.p. Green's 
function. EDF practitioners aiming at functionals
with spectroscopic accuracy believe that a static
approximation, in which the Green's function reduces
to a static density, may be a sound approximation.
This issue asks for a clear answer based on 
comparision of results without uncontrolled
approximations.

\section*{Acknowledgments}

This work is partially supported by the Japanese
Ministry of Education, Culture, Sports, Science and Technology
by Grant-in-Aid for Scientific Research under
the program number (C(2)) 20540277.

\end{document}